\documentclass{sig-alternate-10pt}

\usepackage{times}
\usepackage[utf8]{inputenc}
\usepackage[caption=false,font=footnotesize]{subfig}

\makeatletter
\def\@maketitle{\newpage
\null
\setbox\@acmtitlebox\vbox{%
\baselineskip 20pt
\vskip 1em % Vertical space above title.
\begin{center}
{\ttlfnt \@title\par} % Title set in 18pt Helvetica (Arial) bold size.
\vskip 0em % Vertical space after title.
{\subttlfnt \the\subtitletext\par}\vskip 1em%\fi
{\baselineskip 16pt\aufnt % each author set in \12 pt Arial, in a
\lineskip .5em % tabular environment
\begin{tabular}[t]{c}\@author
\end{tabular}\par}
\vskip 1.5em % Vertical space after author.
\end{center}}
\dimen0=\ht\@acmtitlebox
\unvbox\@acmtitlebox
\ifdim\dimen0<0.0pt\relax\vskip-\dimen0\fi}
\makeatother

\usepackage[citestyle=numeric,firstinits=true,backend=bibtex,minnames=1,maxnames=1,sorting=none,url=false,isbn=false]{biblatex}
\bibliography{biblio}
\AtEveryBibitem{\clearfield{pages}}
\AtEveryBibitem{\ifentrytype{inproceedings}{\clearfield{volume}}{}} % hide volume for non-journal
\AtEveryBibitem{\ifentrytype{inproceedings}{\clearfield{number}}{}} % hide number for non-journal
\AtEveryBibitem{\clearfield{doi}}
\AtEveryBibitem{\ifentrytype{inproceedings}{\clearfield{month}}{}} % hide month for non-journal
\AtEveryBibitem{\ifentrytype{inproceedings}{\clearfield{year}}{}} % hide year for non-journal
\AtEveryBibitem{\clearfield{booktitle}}
\AtEveryBibitem{\clearlist{publisher}}
\AtEveryBibitem{\clearlist{location}} %clearfield clearname clearlist
\renewbibmacro*{issue+date}{%
\setunit{\addcomma\space}% NEW
\iffieldundef{issue}
{\usebibmacro{date}}
{\printfield{issue}%
\setunit*{\addspace}%
\usebibmacro{date}}% NEW
\newunit}

\title{Relaxing constraints in stateful network data plane design}
\author{\rm{
Carmelo Cascone$^{\ddagger}$,
Roberto Bifulco$^{\ast}$,
Salvatore Pontarelli$^{+}$}
\\ 
\rm{
$^{\ddagger}$ Politecnico di Milano,
$^{\ast}$ NEC Laboratories Europe,
$^{+}$ CNIT/Univ. Roma Tor Vergata}
\\
\rm{
carmelo.cascone@polimi.it, roberto.bifulco@neclab.eu,
salvatore.pontarelli@uniroma2.it}}
\begin{document}

\maketitle

\section{Introduction}
% Briefly motivate the need for programming states in the data plane: programmability is good, push applications into the network, offload NFV to hardware.
Modern network devices have to meet stringent performance requirements while providing support for a growing number of use cases and applications. In such a context, a programmable network data plane has emerged as an important feature of modern \emph{forwarding elements}, such as switches and network cards. 
% The need of a network data plane that is programmable has been already recognized by both the research community and the industry.
Bosshart et al. \cite{rmt} introduced RMT, a first example of a high-performance programmable data plane. RMT provides a reconfigurable switching ASIC that can parse and modify arbitrary packet headers in a pipeline of match-action tables (MAT). Interestingly, \cite{rmt} shows that such programmability can be supported with performance comparable to state-of-the-art fixed-function chips: it can process packets at a line rate of 640 Gb/s.

More recently, Sivaraman et al. \cite{domino} presented an abstraction for a switching ASIC, named Banzai, which supports the programming of stateful packet processing functions. The statefullness lays in the ability to create and modify state while processing a packet, enabling the definition of functions that depend on the history of previously received packets.
Such functions enable complex applications such as stateful firewalls, active queue management, scheduling, monitoring, etc.

Banzai extends RMT's MATs by adding stateful actions, named ``atoms''. 
Each atom, as the name suggests, performs state operations atomically. The atomicity is required to guarantee consistency, i.e., read and write operations to an atom's memory area cannot be performed by different packets at the same time. In effect, Banzai requires the serial processing of all the packets. 
This model is convenient since a forwarding element's data plane is already processing packets in a serial manner. 

However, to meet a given performance target, the serial processing model requires the definition of a strict time budget for the processing of each packet. 
For instance, in the case of RMT, the switching ASIC is dimensioned to process 640 Gb/s with minimum size Ethernet packets (64 bytes), which translates to a time budget of 1 ns per packet\footnote{Actually, the minimum size used in the switch may be larger.}. Likewise, the chip clock frequency is dimensioned according to the desired target throughput. In the previous case, a 1 Ghz clock is used to provide the 640 Gb/s. The final outcome is that each atom has to perform state read, modification and write operations in at most 1 ns, i.e., 1 clock cycle.
% to guarantee the 640 Gb/s of throughput. 

Unfortunately, while providing line rate guarantees, Banzai fails to implement more complex functions that require atoms that cannot be executed in the available time budget. E.g., as explained in \cite{domino}, it is not feasible to implement a square root operation in 1 clock cycle at 1 Ghz with a standard 32-nm process.

% translates in a strict constraint to the atom's execution time if one has to guaranteeing line rate performances: in order to prevent consistency of the memory operations, execution of each atom must complete atomically in 1 clock cycle, e.g. in 1 ns at 1 Ghz, such that the switch can process 1 packet every clock cycle, without harming state consistency. Example of an atom is that implementing a counter: for each packet reads a value from the memory, increments it and writes it back, all in 1 clock cycle. 

A solution to this problem could be the partitioning of a complex action's execution over multiple clock cycles. However, if read-modify-write operations happen in two or more distinct clock cycles, the system can easily end up in an inconsistent state. For example, consider an action that implements a packet counter in two cycles, and assume two packets arriving back-to-back. The second packet would cause a read from memory before the processing of the first could write back the incremented counter value, leading to wrong counting. Locking the memory would prevent inconsistencies, but it would also cause the second packet to wait for the first one to update the memory, potentially reducing the switch's throughput. 

% However, while providing line rate guarantees, atoms in Banzai comes with a cost: there will be functions that cannot be implemented as they require more complex instructions to modify memory for which is hard to meet timing in hardware, e.g. as explained in \cite{domino}, it is not feasible to implement a square root operation in 1 clock cycle at 1Ghz with standard 32nm process.
% *** Should we talk about memory interfaces having slower rates (500Mhz), hence requiring more then 1 clock tick to read/write values?

While there seems to be little that could be done to improve on the consistency/throughput trade off, we notice that the design introduced before uses a not very frequent workload to dimension the data plane.
In many common scenarios, network packets have an average size which is bigger than the minimum size.
Therefore, in this work we explore opportunities to relax the constraint of requiring stateful actions to complete execution in 1 clock cycle. Instead, we allow functions that span multiple clock cycles and verify the outcome of such a decision when processing a real traffic trace. In particular, we assess both the risk of harming consistency, in RMT-like architectures, and the potential cost in terms of throughput when introducing locking to still provide consistency guarantees.

\begin{figure*}[]
\centering
\subfloat[][]{
\includegraphics[width=0.285\textwidth]{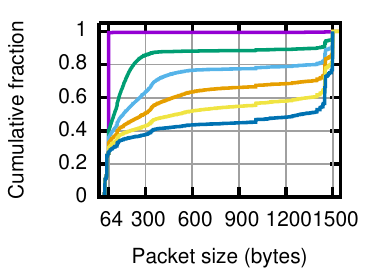}
\label{fig:pkt-size-cdf}
}
\subfloat[][]{
\includegraphics[width=0.285\textwidth]{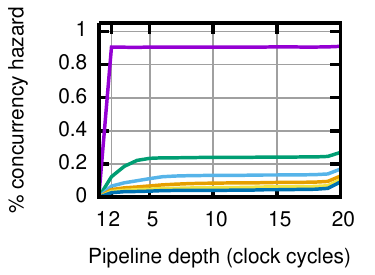}
\label{fig:hazard}
}
\subfloat[][]{
\includegraphics[width=0.38\textwidth]{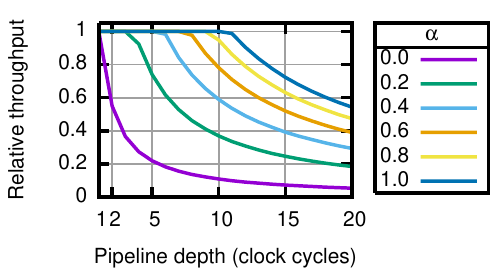}
\label{fig:thrpt}
}
\vspace{-0.3cm}
\caption{Trace-based simulation results}
\label{fig:painchart}
\vspace{-0.3cm}
\end{figure*}

\vspace{-0.3cm}
\section{Contribution}

\noindent
{\bf Background: store-and-forward architectures.}
A design that requires the execution of stateful actions in only 1 clock cycle, derives from the worst case assumption that all packets have minimum size and that they arrive back-to-back, i.e., with no inter-packet gaps. Instead, packets produced by today's application have variable size. E.g., spanning from 64 bytes to 1500 bytes, in a typical case. 

Switches like RMT consume packets in a store-and-forward fashion, i.e., packets are first completely read from input ports, then they are parsed and their \emph{headers} processed by MATs.
With RMT, to sustain a line rate of 640 Gb/s, with a chip clocked at 1 Ghz, we can assume packets are read in chunk of at most 80 bytes (i.e., 80 $\times$ 8 bit $\times$ 1 Ghz $=$ 640 Gb/s). Consequently, it will take 1 clock cycle to read packets with minimum size $\le 80$ bytes, while it will take more cycles to read longer packets, e.g., 19 for 1500 bytes. Still, the switch's pipeline can process the parsed headers in one clock cycle, independently from its size. The result is that, even when all packets arrive back-to-back, the variability of the packet size will cause the pipeline to experience one or more idle cycles. Can we exploit this consideration to relax the atomicity constraint?

\noindent
{\bf Trace-based simulations.}
We run simulations using real traffic traces in order to (i) assess the risk of harming consistency when using stateful functions that span many clock cycles, and to (ii) evaluate the throughput when introducing locking to provide consistency guarantees. We used packet traces from a US backbone link collected in 2015 \cite{caida2015}. We accelerate the trace to drive 100\% utilization, removing inter-packet gaps. 
Figure~\ref{fig:pkt-size-cdf} shows the cumulative distribution function (CDF) of the packet size found in the trace. 
We further define $\alpha$ as a parameter to modify the size of each packet in order to gradually approximate the worst case. When $\alpha = 0$ all packets have minimum size, i.e., our worst case. Instead, with $\alpha = 1$, we use the trace's orginal packet size distribution. Finally, we model the packet size modification function such that intermediate values of $\alpha$ represent realistic cases, e.g. when $\alpha = 0.2$ the CDF is similar to that found in Facebook's datacenter \cite{fbdc}.
% roughly 50\% of packets have size between 1250 and 1500 bytes, meaning that, for about 50\% of time, there will be a gap of 16-18 clock cycles between consecutive packets. 
% Similarly, when $\alpha = 1$ only 30\% of packets have minimum size.

Assuming a stateful action implemented as a pipeline of many instructions, where the first reads from the memory and the last writes back, we define as ``concurrency hazard'' the event in which the first instruction of the action pipeline processes a packet, while another one is currently travelling in the same pipeline. Figure~\ref{fig:hazard} shows the percentage of time such an event occurs with the chosen traffic trace, while varying the pipeline depth for different values of $\alpha$. Clearly when $\alpha=0$ the risk of concurrency hazard is maximum already with pipelines of 2 instructions, since we have exactly one packet per clock cycle to process. However, when $\alpha$ tends to 1, the hazard grows slowly. When $\alpha=1$, the risk of concurrency hazard is lower than 5\% with action pipelines up to 19 clock cycles. This result suggests that a locking strategy to prevent inconsistencies could be applicable, with marginal harm to the throughput, if any. 

We consider now the case of a trivial locking strategy that allows only 1 packet at a time in the action pipeline. Clearly, locking affects throughput, but most importantly voids the guarantee of deterministic performance. Figure~\ref{fig:thrpt} shows the throughput is not reduced with actions pipelines up to 10 clock cycles when $\alpha = 1$, i.e. when processing traffic with characteristics similar to a backbone link. In fact, larger packets provide the pipeline with additional time to process smaller packets that may have been queueing because of the locking. Conversely, maximum throughput is guaranteed only up to 3 clock cycles when $\alpha = 0.2$, i.e. with traffic similar to that found in Facebook's datacenters.

% \section{Discussion}
\vspace{-0.3cm}
\section{Future work}
Our early results suggest that there could be several cases in which switches may apply operations that require multiple clock cycles per packet, while still maintaining line-rate throughput and consistency. Therefore, blocking architectures could be a viable option for implementing complex stateful actions in forwarding elements.

Our work is however in an early stage, despite the need to verify our assumptions against a larger number of traffic traces from different scenarios, there are still a number of unexplored issues.
For instance, in several cases packets belonging to different flows do not read/write each others' states. A characteristic that could be exploited to implement a smarter locking scheme. Furthermore, packet processing actions may be different in length, with some taking longer than others. If multi-cycle actions are uncommon, there could be the possibility to implement some very complex actions together with a larger set of simpler ones.

Exploring the above points and providing an effective hardware implementation that can leverage the corresponding findings is part of our future work.

% \section{Future work}
% to demonstrate that the observation holds in a vast number of real world cases; to provide a cost effective hardware implementation

%\bibliographystyle{abbrv}
%\bibliography{biblio}
%\AtNextBibliography{\small}
{%\scriptsize
\vspace{-3mm}
\printbibliography
}

\end{document}